\documentclass{aa}     
\usepackage[dvips]{graphics}
\usepackage{latexsym}
\newcommand{\cmc} {cm$^{-3}$}

\newcommand{\kms} {km s$^{-1}$}

\newcommand{\Myr} {M$_\odot$ yr$^{-1}$}
\newcommand{\um} {$\mu$m}
\newcommand{\mic} {$\mu$m}

\newcommand{\sbu} { erg cm$^{-2}$ s$^{-1}$ sr$^{-1}$}
\newcommand{\Lsun} {L$_\odot$}
\newcommand{\Msun} {M$_\odot$}
\newcommand{\Filyc} {$\Phi_i$}

\newcommand{\Teff} {T$_\star$}
\newcommand{\Tstar} {T$_\star$}
\newcommand{\Lstar} {L$_\star$}
\newcommand{\Mstar} {M$_\star$}
\newcommand{\Hmol} {H$_2$}
\newcommand{\brg} {Br$\gamma$}

\newcommand{\simless}{\mathbin{\lower 3pt\hbox
      {$\rlap{\raise 5pt\hbox{$\char'074$}}\mathchar"7218$}}} 
\newcommand{\simgreat}{\mathbin{\lower 3pt\hbox
     {$\rlap{\raise 5pt\hbox{$\char'076$}}\mathchar"7218$}}} 

\begin{document}
 
\thesaurus{}
 
\title{A NEAR-INFRARED STUDY OF THE PLANETARY NEBULA NGC~2346
\thanks{Based on observations obtained at the TIRGO telescope}} 
 
\author { B. Vicini\inst{1}, A. Natta\inst{2}, 
          A. Marconi\inst{2}, L. Testi\inst{3},
          D. Hollenbach\inst{4} and B.T. Draine\inst{5}}
 
\institute{
    Dipartimento di Astronomia e Scienza dello Spazio, Universit\`a
    degli Studi di Firenze, Largo E.Fermi 5, I-50125 Firenze, Italy
   \and
    Osservatorio Astrofisico di Arcetri, Largo E.Fermi 5,
    I-50125 Firenze, Italy
   \and
    Division of Physics Mathematics and Astronomy, Caltech,
    MS 105-24, Pasadena, CA 91125, USA
    \and
    NASA Ames Research Center, MS 245-3, Moffett Field, CA 94035, USA
    \and
    Princeton University Observatory, Peyton Hall, Princeton, NJ 08544, USA
           }
 
\offprints{natta@arcetri.astro.it}
\date{Received ...; accepted ...   }
 
\titlerunning{NGC~2346}
\authorrunning {B. Vicini et al.}

\maketitle

\begin{abstract}
This paper presents new near-infrared 
observations of the planetary nebula NGC~2346.
The data include 
a broad K--band image, an image in the \Hmol\ vibrationally excited
1-0S(1) line  and K band slit spectra
at three positions in the nebula.
In the \Hmol\ 1-0S(1) line, the nebula is characterized by a central,
bright torus, surrounded by weaker emission with a typical
butterfly shape, as seen in H$\alpha$ and CO lines.
The K band spectra show 11 \Hmol\ lines with excitation energies from
6150 to 12552 K.
The \Hmol\ data have been compared to the predictions of models
which follow the evolution with time of the \Hmol\ emission in PNe 
of different core mass and
shell properties (Natta \& Hollenbach 1998). These
 models compute the emission originating
in the photodissociation region (PDR) created at the inner edge of
the neutral shell by the UV radiation of the
central core, as well as
the emission in the shock associated to the expansion of the shell
inside the precursor red-giant wind. 
In NGC~2346, a PDR origin of the \Hmol\ emission in a low-density
molecular shell ($n\simless 10^4$ cm$^{-3}$) is indicated. At these low
densities, time-dependent
\Hmol\ chemistry and X-ray heating of the neutral gas enhance
the predicted  PDR \Hmol\ line intensity by  large factors.

\end{abstract}
 
\section{Introduction}
 
NGC~2346 is a much studied bipolar planetary nebula at a distance
$D\sim 800$ pc (Acker et al. 1992)\footnote{ After this work was completed,
a new determination of the
distance ($D$=690 pc) was quoted by Terzian (1997). It is close enough to the
value we adopt in this paper that none of the conclusions needed to be changed.}.
 At its center lies
a binary system, formed by a main-sequence star of spectral type A5V,
with mass $\sim 1.8$ \Msun, temperature $\sim 8000$ K and luminosity
$\sim 14$ \Lsun (M\'endez and Niemela 1981; Walsh 1983) 
and hot star, not detected in the visual,  with \Teff$\sim 130000$
K (M\'endez 1978), which excites the nebula. Its luminosity is very
uncertain, as we will discuss in \S 4. Estimates in the
literature give \Lstar$\sim 17-90$ \Lsun (M\'endez 1978;
Calvet and Peimbert 1983).

In the optical, the nebula has a butterfly shape (Balick 1987;
Walsh et al. 1991), with well
developed bipolar lobes and a bright torus which surrounds the central star.
The temperature and density of the ionized gas in the torus have
been estimated to be $T\sim 12000$ K and $n_e\simless 700$ \cmc, respectively
(Liu et al. 1995; McKenna \&  Keenan 1996).
The nebula contains a large amount of material in the form of molecular
gas, as revealed by the CO observations of Knapp (1986), Huggins \& 
Healy (1986), Healy \& Huggins (1988). Bachiller et al. (1989) have mapped
the entire nebula in the two CO lines J=1-0 and J=2-1; the morphology of
the molecular gas follows very closely that of the ionized gas,
showing a clumpy, inhomogeneous torus, 
tilted with respect to the line of sight by about 56$^o$,
which is expanding outward.
Scaled to our
adopted distance $D=800$ pc, the torus has a radius of $\sim$0.05 pc, and
mass $\sim$ 0.26 \Msun, much larger than the mass of ionized gas
($\sim$0.01 \Msun, Walsh 1983).
The 
radial velocity of the most intense CO condensations is of the order
of 15--35 \kms, which results in a dynamical age of about 2500 yr
(but see also Walsh et al. 1991).

NGC~2346 is a Type I nebula, originated by a massive
progenitor (Calvet and Peimbert 1983).
As many PNe of similar type, NGC~2346 is  detected in the
vibrationally excited lines of \Hmol (Webster et al. 1988).
Zuckerman \& Gatley (1988) have mapped the nebula in the 1-0S(1)
line  at 2.12 \um\ using a single-beam 12 arcsec spectrometer
with  resolution $\sim$ 200. The
morphology of the nebula in this line is again very similar to the 
morphology observed in the optical lines and in CO. This result
was confirmed more recently by the images obtained in the same line
with much better spatial resolution (about 1-2 arcsec) by Kastner et al. (1994)
and Latter et al. (1995).

The excitation mechanism of the vibrationally excited \Hmol\ lines in this,
as in other PNe, is still uncertain. Zuckerman \& Gatley (1988)
discuss the possibility that they form in a shock driven by the fast wind 
emitted by the central star.
Kastner et al. (1994) surveyed a sample of bipolar planetary nebulae
(including NGC~2346); they conclude that the \Hmol\ emission very
likely originates
in  thermally excited (possibly shocked) molecular gas.
Recently,  Natta \& Hollenbach (1998;
hereafter NH98) have computed theoretical
models of the evolution of PN shells and predicted, among others, the
intensity of the most commonly observed \Hmol\ vibrationally excited
lines (namely, the 1-0S(1) at 2.12 \um\ and the 2-1S(1) at 2.25\um). 
They consider     the emission of the photodissociation region
(PDR) formed by the UV photons emitted by the
central star impinging on the shell,
including in the calculations time-dependent \Hmol\ chemistry
and the effects of the soft X-ray radiation emitted by the central star,
which are important in sources like NGC~2346 where \Tstar$\simgreat 10^5$ K.
NH98 compute also  the emission of the shocked gas
at the interface between the shell and the  wind ejected by the
central star in its previous  red giant phase.
They point out that both mechanisms (PDR and shocks)
can produce lines of similar
intensity, with reasonable values of the model parameters.

The PN properties that determine the intensity of the \Hmol\ lines are
very different in the two cases. As discussed
in NH98, if the emission is produced in the
warm, neutral PDR gas, the line intensity depends mostly on the stellar
radiation field which reaches the shell and, to a lower degree, on
the density of the neutral gas itself. If the emission is produced in
the shocked gas, then the line intensity does not depend directly on the
properties of the central star or of the PN shell, but only on
the shock velocity and on the rate of mass-loss of the precursor red-giant.
It is therefore clear that, before attributing any diagnostic capability to the
\Hmol\ lines, we need to understand which of the possible excitation mechanisms
dominate the PN emission.

This paper is a first attempt to understand the \Hmol\ emission of
a well-studied PN in a quantitative way, i.e., by comparing
the observations 
to detailed models of PDR and
shock emission, such as those discussed in NH98.
To this purpose, we have collected
new near-IR broad and narrow-band images of NGC~2346  as well as 
K band spectra with resolution $\sim$ 1000.
These observations are described in \S 2. The results are 
described in \S 3 and  compared to
the predictions of PDR and shock models in \S 4. 
A discussion of the results follows in \S 5; \S 6 summarizes the main
conclusions of the paper.

\section {Observations}

\subsection {ARNICA Observations}

NGC~2346 was
observed during two observing runs in January 1996 
using ARNICA (ARcetri Near Infrared CAmera)
mounted on the 1.5m TIRGO\footnote{The
  TIRGO telescope is operated by the C.A.I.S.M.I.-C.N.R
   Firenze, Italy} telescope. ARNICA is equipped with a 256x256 NICMOS3
array, the pixel size with the optics used
at TIRGO is $0.96^{\prime\prime}$; for a complete
description of  instrument  performances, see Lisi et
al.~(\cite{Lea96}) and Hunt et al.~(\cite{Hea96}).
Images were obtained in the  K broad-band filter 
(centered at 2.2 \um) and in  a
narrow-band filter centered 
on the 2.12 \Hmol\ 1-0S(1) line 
($\Delta\lambda/\lambda\sim 1\%$, Vanzi et al.~\cite{VGCT97}
).
The seeing was approximately 2-3\arcsec  and
the observed field was 
$\sim 2^\prime\times 2^\prime$,
covering all the nebula.

Data reduction was carried out
using the 
IRAF\footnote{IRAF is made available to the astronomical
community by the National Optical Astronomy Observatories,
which are operated by AURA, Inc., under contract with the U.S.
National Science Foundation} and ARNICA (Hunt et al. 1994) software packages.
Photometric calibration in the  K band was
performed by observing  the photometric standards
of the FS14 group from the list of Hunt et al.~(\cite{Hea97}).
The quality of the night was rather poor, and the calibration
accuracy is estimated to be  $\sim 15\%$.

The image in the \Hmol\ 1-0S(1) line has been calibrated using the 5 brightest 
(unsaturated) stars
in the ARNICA images, under the assumption that for each star the flux
density
measured in the line filter was equal to the flux density measured in the K 
band. Integrated line fluxes on the nebula were then obtained multiplying the
flux density by the bandwith of the narrowband filter (Vanzi et al. 1998).
The accuracy is $\sim$15\%.

\subsection {LONGSP Observations}

 K (2.2 \um )
band spectra of NGC~2346 were obtained using the
LonGSp (Longslit Gornergrat Spectrometer) spectrometer mounted
at the Cassegrain focus on the TIRGO telescope.
The spectrometer is equipped with cooled reflective optics and
grating in Littrow configuration. The detector is a
256$\times$256 engineering grade
NICMOS3 array (for detector performances see Vanzi et al.~\cite{VMG95}).
The pixel sizes are 11.5 \AA\ (first order) and 1\farcs73
in the dispersion and slit directions, respectively.
LONGSP operates in the range 0.9-2.5 \mic\ achieving
a spectral resolution at first order of 
$\lambda/\Delta\lambda\simeq$950 in K.
For a more comprehensive description of the
instrument, refer to Vanzi et al.~(\cite{Vea97}).
 
Observations were conducted in 
March 1998 under non-photometric
conditions. The slit
had dimensions
3\farcs5$\times$70\arcsec\ and was oriented N-S. The seeing
during the observations was in the range 2\arcsec--4\arcsec.
NGC~2346 was observed at three slit positions labeled
as E, S, W and shown in Fig.~\ref{fig:imh2} superimposed on  the
image in the \Hmol\ 1-0S(1) line.
Position E and W  are centered on the peaks of the line emission
located east and west of the central star, respectively
(see Zuckerman and Gatley 1988).
Position S  is centered on the star.
At each grating position we
performed 5 ABBA cycles (A=on source, B=on sky) with an on-chip
integration time of 60 sec, for a total of 10 min integration on source.

Data reduction was performed with the ESO package MIDAS, within
the context IRSPEC, modified to take into account LonGSp
instrumental characteristics.
The frames were corrected for bad pixels,
flat-fielded, sky subtracted and wavelength
calibrated using the OH sky lines present
in all the frames (Oliva \& Origlia~\cite{OO92}).
After  direct subtraction, sky removal
was optimized by minimizing the standard deviation
in selected areas where the OH sky lines were poorly
subtracted but no object emission was present.
The wavelength calibration was performed to better than 1/5 of
a pixel ($\simeq$2\AA).
 The spectra were then corrected for telluric absorption
by dividing the spectra by the
spectrum of the A star 
BS~2714
after removing its photospheric features (mainly Br$\gamma$).
For more details on LonGSp data reduction, see Vanzi et al. \cite{Vea97}.

Flux calibration of the spectra 
was achieved by rescaling the
observed flux distribution along the slit to match that obtained
from the ARNICA image in the 1-0S(1) line at the positions of the slits.

\section {Results}

\begin{figure}
\resizebox{\hsize}{!}{\includegraphics{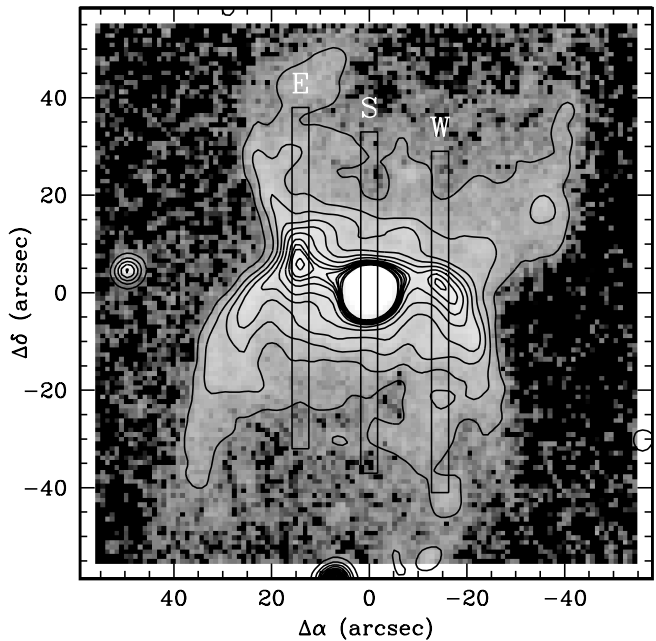}}
\caption{  Image in the H$_2$ 1-0S(1) line.  The contours  range from
1 to 14 in steps of $1.3 \times 10^{-5}$ erg cm${-2}$ s$^{-1}$
sr$^{-1}$.
The three positions of the
slit used to obtain the K-band spectra are shown by the black
boxes and are marked as E, S and W, respectively. The continuum (not subtracted 
from the image) is confined to a region of $\sim 5$\arcsec\
radius centered on the 
star.}
\label{fig:imh2}
\end{figure}

The image in the  
\Hmol\ 1-0S(1)  line
is shown in 
Fig.~\ref{fig:imh2}.
As verified with the spectra, the continuum emission is everywhere
negligible but at the position of the central star, i.e. in
a 5\arcsec\ radius region centered on the star.



The \Hmol\ 1-0S(1) image of Fig.~\ref{fig:imh2} shows the well-known NGC~2346 morphology, with
a bright central region of size $\sim$ 50\arcsec$\times$20\arcsec\
and two very extended 
lobes of weaker emission (Kastner et al. 1994).
The central region  has two peaks of emission,  to the east and  west
of the star and matches well the bright torus, tilted with respect to
the line of sight, seen in optical tracers and in CO (Walsh 1983;
Bachiller et al. 1989). 
The H$_2$ 1-0S(1) intensity is  $\sim 1.3\times 10^{-4}$ \sbu
on both peaks.
The total luminosity of the nebula in this line is about 0.06 \Lsun (for $D=800$
pc), of which about 48\% is contributed by the torus. The average line
intensity over the torus (defined as the central region of size
20\arcsec$\times$50\arcsec) is  $\sim 6\times 10^{-5}$ \sbu.
These numbers are very similar (within 10-20\%) to those derived by
Zuckerman and Gatley (1988).

\begin{figure}
\resizebox{\hsize}{!}{\includegraphics{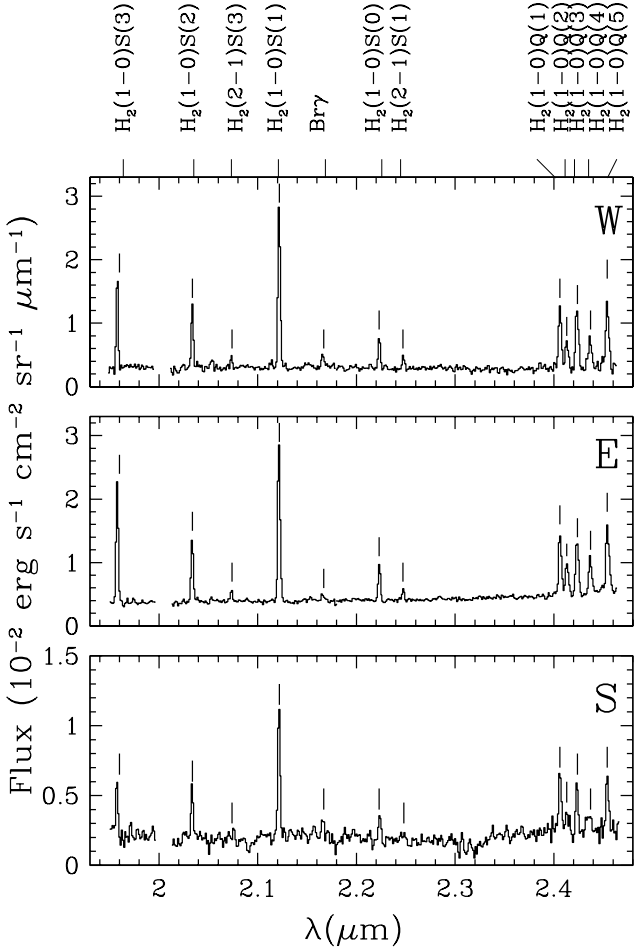}}
\caption{K-band  spectra in the W, E and S positions. The lines have been
averaged over a slit section of 
$\pm$10\arcsec\ centered on the peak of the emission
(the torus mid-plane). The small gap at $\sim$2--2.04 \um\ corresponds to a region
of very poor atmospheric transmission.}

\label{fig:spec}
\end{figure}

The K band spectra in the three positions E, S and W are shown 
in Fig.~\ref{fig:spec}. The spectra have been averaged
over a region of $\sim$20\arcsec\ 
(12 pix) along the slit centered on the torus midplane.
The line intensities 
are given in Table 1 , which gives
in Column 1
the  line identification, 
in Column 2 the wavelength of the line, in Column 3
the intensity in the W position, in Column 4 that on the E
position, in Column 5 that in the S position of the slit.
The   lines are normalized
to the 1-0S(1) line  set equal to 100; 
the intensity of the 1-0S(1) line
is given in the Table's note.
Typical uncertainties  on the line ratios are  $\sim$10\% for
ratios $>$50, and
$\sim$30\%
for the others. 
Lines whose intensity is particularly uncertain are marked with
a semicolon.

\begin{table}
\begin{center}
\caption{K-Band Spectrum of the Central Torus}
\vskip 0.1cm
\vbox{\hskip -8mm
\begin{tabular}{lcccc}
\hline\hline
\medskip
Line&  $\lambda$& W& E& S\\
    &  ($\mu$m)&  &  &  \\     
\hline
\smallskip
1-0S(3)& 1.958& 50 & 68& 43\\
1-0S(2)& 2.034& 36& 37& 39\\
He I&    2.058& 6:& $<$2& $<$6\\
2-1S(3)& 2.073& 6.2& 5.9& 9:\\
1-0S(1)& 2.122& 100& 100& 100\\
\smallskip
2-1S(2)& 2.154& $<$4&$<$4&$<$11\\
Br$\gamma$&2.166&8.7&4.7 &$<$16\\
3-2S(3)& 2.201& $<$4& $<$4& --\\
1-0S(0)& 2.223& 20& 21& 23\\
2-1S(1)& 2.248& 7.0& 6.7& $<$9\\
\smallskip
3-2S(2)& 2.287& $<$4& $<$4& --\\
2-1S(0)& 2.355& $<$4& $<$4& --\\
3-2S(1)& 2.386& $<$4& $<$4& --\\
1-0Q(1)& 2.406& 44 & 44& 49\\
1-0Q(2)& 2.415& 20& 25& $<$20\\
\smallskip
1-0Q(3)& 2.424& 46& 41& 35\\
1-0Q(4)& 2.437& 24& 26& $<$20\\
1-0Q(5)& 2.454& 51& 55& 43\\
\hline\hline
\end{tabular}}
\end{center}
Note: 1-0S(1) intensity 100 corresponds to
$7\pm 2\times 10^{-5}$ erg cm$^{-2}$s$^{-1}$ sr$^{-1}$ in the W position,
$9\pm 3\times 10^{-5}$ erg cm$^{-2}$s$^{-1}$ sr$^{-1}$ in the E position 
and $3\pm 1\times 10^{-5}$ erg cm$^{-2}$s$^{-1}$ sr$^{-1}$ in the S 
position.
The spectra have been averaged over a slit portion $\pm$10\arcsec\
centred on the torus midplane.\\
\label{table:spectra}
\end{table}

%
%

\begin{figure}
\resizebox{\hsize}{!}{\includegraphics{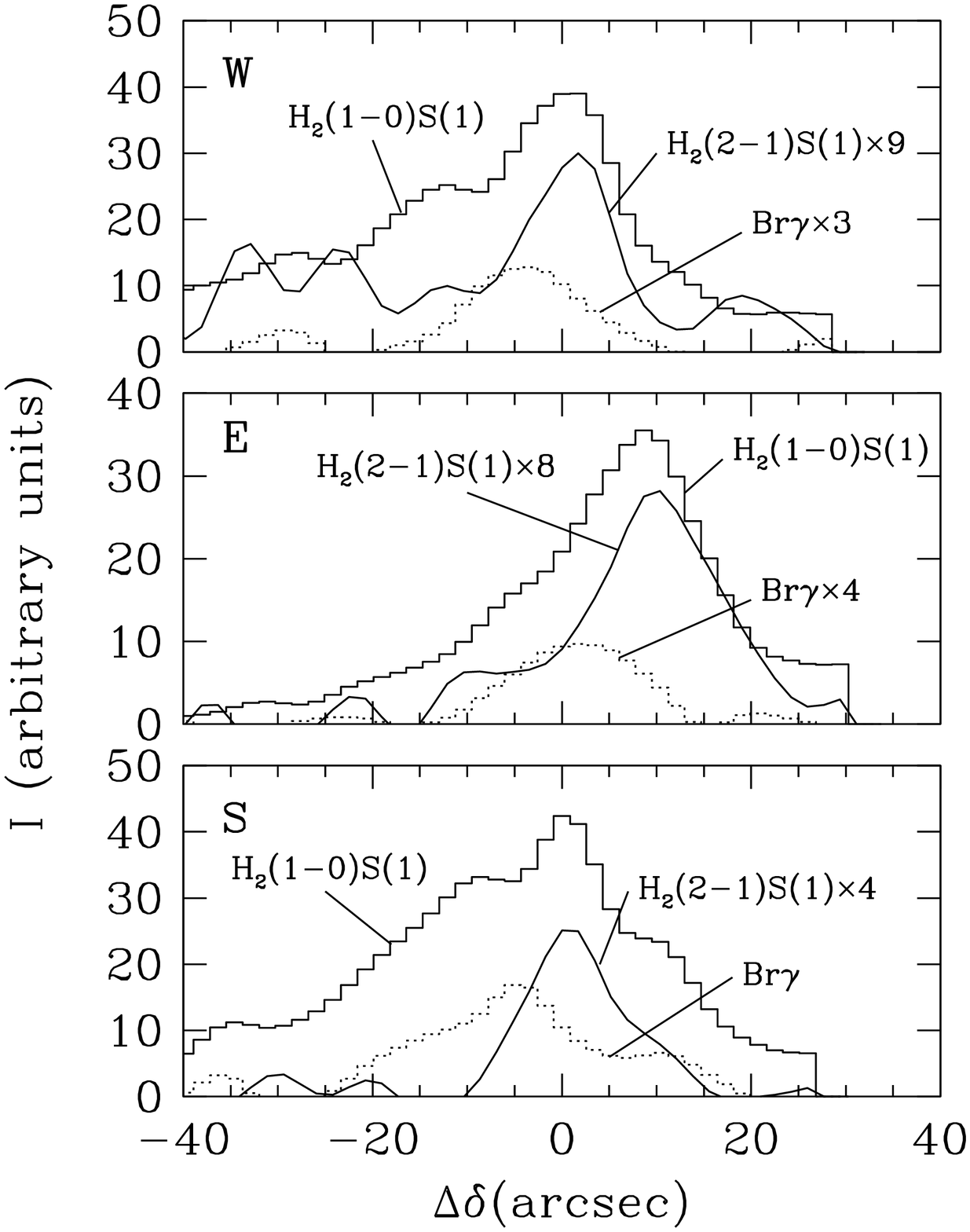}}
\caption{ Top panel:  variation with declination offset of the intensity
of the H$_2$ 1-0S(1) and 2-1S(1) lines
and of \brg\ (dotted curve) for the slit position E.
Middle panel: same for slit position W. Bottom panel: same for slit position S.
All intensity profiles have been smoothed over three
pixels. The offset is measured with respect to the central star.}
\label{fig:cuts}
\end{figure}

The variations
in intensity along the
slit positions E and W and S of  the \Hmol\ 1-0S(1) and 2-1S(1) lines
and of \brg\
is shown in Fig.~\ref{fig:cuts}.
The intensity profiles show that the \Hmol\ emission in the 1-0S(1)
line peaks in the torus (i.e., for $|\delta|\simless$ 10\arcsec)
and has extended emission with intensity
that declines more sharply toward the north than toward
the south, possibly due to the tilt in the plane of the
sky. Some of the inhomogeneities seen in the two-dimensional
image can also be seen in these profiles.
The ratio of the 2-1S(1) over the 1-0S(1) intensity 
varies between  about 0.1  in the W and E  peaks of 
emission
to about 0.04 in the condensation detected in the W position
$\sim$15\arcsec\ south of the midplane. 
Another interesting result which emerges from the spectra and the intensity
profiles is that the emission in \brg\ 
extends over
the whole central torus. However, 
\brg\ is quite weak, $\sim$10\% of the 1-0S(1).

\section {Comparison of the \Hmol\ spectrum with models}

The \Hmol\ emission in the 1-0S(1) line 
of  the {\it central torus} of NGC~2346 is
compared in  this section to the 
predictions of the NH98 theoretical models.

The NH98 models compute the emission 
expected from tori (or shells  or  clumps)
exposed to the radiation field of the hot central star and expanding
inside the remnant of the wind ejected by the central star
in  its previous red giant  phase.
As the central star evolves initially at constant
luminosity  toward higher effective temperatures
and then along the white dwarf cooling track, the torus expands
radially 
with roughly constant velocity and decreasing density.
The models consider the effects of UV and soft X-ray radiation
on the neutral gas and follow the time dependent chemistry for \Hmol,
solving for the chemical and temperature structure 
and the emergent spectrum (in the following, the PDR spectrum)
of the evolving torus.

The torus expands  inside the material ejected by the star in its
previous phase as  red giant. Since its velocity is
larger than that of
the wind itself, it gives origin to a shock which heats and compresses the gas,
which  then emits intense lines of vibrationally excited \Hmol.
NH98 consider the emission of J shocks, since the importance of magnetic
field in PNe is not clear and, in  this range of shock velocities, the
\Hmol\ line intensity is maximum in J shocks.

\subsection {PDR Models}

\begin{figure}
\resizebox{\hsize}{!}{\includegraphics{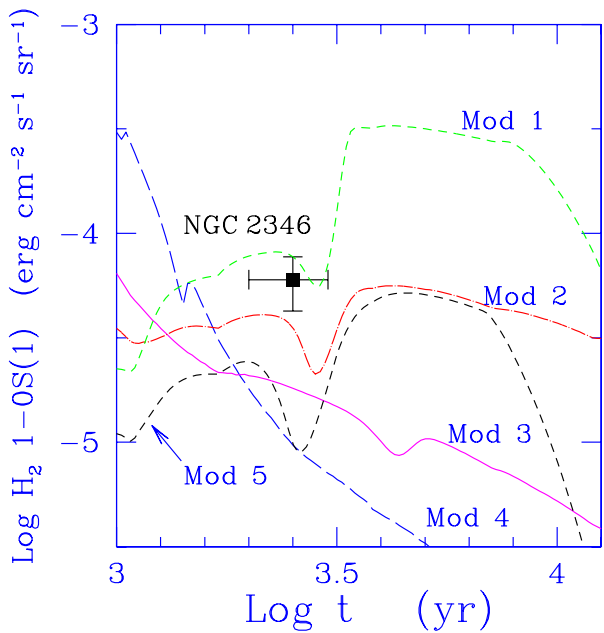}}
\caption{ Model predictions for the \Hmol\ 1-0S(1) line intensity
as
function of the age of the nebula. 
Model parameters are given in Table~2.
All the models have a neutral gas density decreasing as $t^{-2}$ as
the torus expands with constant velocity $v=25$\kms.
The square shows the value of the observed \Hmol\  1-0S(1)
intensity averaged over the central torus.}
\label{fig:PDR}
\end{figure}

The NH98 PDR models require the specification of  a number of parameters.
We fix the mass of
the central core, which determines the time scale of the evolution of
the stellar radiation field,  to be about 0.7\Msun (Calvet and
Peimbert 1983; Bachiller et al. 1989). 
For a core mass \Mstar=0.7 \Msun, NH98 show the results of only one model, 
which has a neutral gas density at the time
t=2500 yr (roughly the age of NGC~2346)
$n_0\sim 7\times 10^5$ \cmc (note that in NH98 the density
varies with time as $n=n_0 (t/t_0)^{-2}$). 
We used the NH98 code to 
compute a number of additional models, varying the  $t=2500$ yr density
$n_0$
from $7\times 10^3$ \cmc (Model 1) to $2.1\times 10^4$ \cmc (Model 2),
$7\times 10^4$ \cmc (Model 3) and $7\times 10^5$ \cmc (Model 4). 
The model parameters are summarized in
Table~2.
In all cases, He/H=0.13, C/H=$5.3\times 10^{-4}$, O/H=$4.6\times 10^{-4}$.
The results are shown in Fig.~\ref{fig:PDR}, which plots 
the intensity of the \Hmol\ 1-0S(1) line  as a function of time.

\begin{table*}
\begin{flushleft}
\caption{ Model Parameters}
\vskip 0.1cm
\begin{tabular}{lccccc}
\hline\hline
\medskip
& Mod 1& Mod 2& Mod 3& Mod 4 & Mod 5\\
\hline
\smallskip
$M_\star/M_\odot$& 0.7& 0.7& 0.7 & 0.7& (a)\\
$v$ (km s$^{-1}$)& 25& 25& 25& 25& 25\\
$n_0$ (cm$^{-3}$)& $7\times 10^3$& $2.1\times 10^4$& $7\times 10^4$& $7\times 10^5$ & $7\times 10^3$\\
\hline
\hline
\end{tabular}\\
(a): Mod 4 assumes a luminosity of the central core equal to \\
\Lstar(\Mstar=0.7)/5 at all times.
\label{table:models}
\end{flushleft}
\end{table*}
 
The square in Fig.~\ref{fig:PDR}
shows the  observed  intensity of the 1-0S(1) line
averaged over the torus.
We have estimated the age of NGC~2346 by taking the separation of the two 
1-0S(1) peaks
(about 30 \arcsec)
at  $D=800$ pc and an expansion velocity of $\sim$25 \kms (as in the NH98
models), consistent with the observed range (15--35 \kms) of CO
expansion velocity (Bachiller et al. 1989).
The result is a  dynamical age of $\sim$2500 yr.
The uncertainty on the age is certainly large, but difficult to estimate.
We plot in Fig.~\ref{fig:PDR} an error bar corresponding to an uncertainty
of $\pm$500 yr.
The observed intensity of the 1-0S(1) line
is reproduced quite well by the low-density
models, especially if we take into consideration
the large uncertainties that affect the models as well as the age estimate of
the nebula itself. 

One problem that arises immediately with these models has to do with the
estimated luminosity of the central star.
At $t\sim 2500$ yr,
a core of 0.7 \Msun\ has already reached the white dwarf
cooling track; it has a luminosity of about 250 \Lsun, an
effective temperature \Tstar$\sim 1.5\times 10^5$ K and
a number of ionizing photons \Filyc$\sim 3\times 10^{45}$ photons s$^{-1}$
(Bl\"ocker 1995). While the effective temperature is roughly in
agreement with the  Zanstra HeII temperature (M\'endez 1978),
the luminosity is  significantly 
larger than the   values 17--90 \Lsun\ quoted in the literature.
However,
we are somewhat suspicious of those very low values. 
The luminosity derived from the HeII $\lambda$4685\AA\ intensity
by M\'endez (1978) is a lower limit \Lstar$>$43 \Lsun (for $D$=800 pc) 
and is very
sensitive to the extinction.  
The number
of ionizing photons we derive from the observed radio flux at 6 cm
(86 mJy; Milne and Aller 1975) and from the total H$\alpha$ flux (Walsh 1983) is at least 
$2-4 \times 10^{45}$ photons s$^{-1}$ (assuming  no escape of ionizing photons
and an average optical depth in H$\alpha \sim 0.7$), consistent
with \Lstar$\sim$250 \Lsun, but not with lower
values of \Lstar
(see Fig.~3 of NH98).
If \Lstar=250 \Lsun,
the non-detection of the white dwarf star in the visual is not surprising;
assuming, for simplicity, that the A star and the white dwarf spectrum can
be represented by black-bodies at 8500 K and $1.5\times 10^5$K, having
luminosities of 15 and 250 \Lsun, respectively,
we find that the white dwarf  is a factor 54 weaker  than  the A-type
star at 5500 \AA, a factor of 10 at 3000\AA\ and that the two stars become comparable
only at $\sim$2000\AA. 
This last is consistent with the UV excess (with respect to
the flux expected for the A star) measured by the ultraviolet satellite
ANS and reported by M\'endez (1978).
All together, we suspect that the white dwarf luminosity is
roughly of the
order of 250 \Lsun.
In any case, we have also computed a model where we 
have artificially reduced the stellar luminosity 
by a factor 5 
at all times; the density of this model (Model 5),
 shown as a dashed curve in Fig.~\ref{fig:PDR},
is $n_0=7\times 10^3$ \cmc. 
The predicted line
luminosity scales approximately with the luminosity of the
central core.  A value \Lstar=50 \Lsun\ (although not predicted
by any evolutionary track)
is still roughly consistent with the observed
line intensity, especially if we consider the uncertainty on the PN
age estimate. However, this model predicts an intensity lower than observed
(by a factor 3--10) for all the \Hmol\ lines we  measured.

The best fit to the \Hmol\ 1-0S(1) observations is provided
by models with low density ($n_0\sim 1-3\times 10^4$ \cmc),
in good agreement with the low electron density ($\simless 10^3$ \cmc)
derived for the ionized part of the nebula (Liu et al. 1995; McKenna \& Keenan 1996).
Assuming pressure equilibrium between the ionized and the neutral gas,
and a PDR temperature $\sim$500 K, we expect a neutral density
about 40 times the electron density.
The low density we require is in rough agreement with the  Bachiller
et al. (1989) estimate that  the neutral density is $few \times 10^3$ \cmc.

In these low-density models, at $t\sim 2500$ yr, the 1-0S(1) emission is mostly due
to collisionally excited \Hmol, kept warm by the heating of the soft X-rays
emitted by the central core. As discussed in NH98, X-rays determine
the chemical and physical evolution of the neutral gas around
high-mass PN cores, after a short initial phase  (about 1000 yr
for a 0.7 \Msun\ core) where UV photons dominate.  If the X-rays effects are
neglected, PDR models predict a much lower intensity of the \Hmol\
molecular lines.

Time-dependent effects in the
\Hmol\ chemistry are important. For \Mstar=0.7 \Msun\ and
$n\propto t^{-2}$, $t\simgreat 10^3$ yr,
the mass of ionized gas increases with time. In these conditions, at
each time step a new layer of molecular gas is exposed to the X-ray
heating radiation, as 
\Hmol\ molecules are advected from deep in the PDR slab toward the
irradiated surface.
Therefore, compared to
the predictions of equilibrium calculations,
a larger amount of hot molecular gas 
is formed,  which emits stronger \Hmol\ vibrationally excited lines.
This effect, discussed in detail in NH98, is larger in models with
lower density, so that  the 1-0S(1) intensity
is higher in models with lower $n$. The opposite is true in models
where the \Hmol\
chemistry is treated under the assumption of stationary equilibrium.
These models  predict at $t\sim 2500$ yr
a 
1-0S(1) line about 7-10 times weaker  (for $n_0\leq 2.1\times 10^4$ cm$^{-3}$),
mostly due to
fluorescence in \Hmol\ pumped by  UV photons,
and lower in models with lower $n$.

The predicted intensities of all the \Hmol\ observed lines have  been computed
using a code
 which calculates the level population 
for the $N$=299 bound states with rotational quantum number
$J\leq$29 of the \Hmol\ molecule. 
The code include the effects of UV pumping by an
external radiation field as well as collisions with
H, \Hmol, He, electrons and protons (Draine \& Bertoldi 1996). We
have used as input the physical conditions (namely,
the radiation field at the inner edge of the PDR and the run
with the depth in the PDR
of temperature and fractional abundances of H, \Hmol, He, electrons
and protons) computed with the NH98 code for $t=2500$ yr.

\begin{figure}
\resizebox{\hsize}{!} {\includegraphics{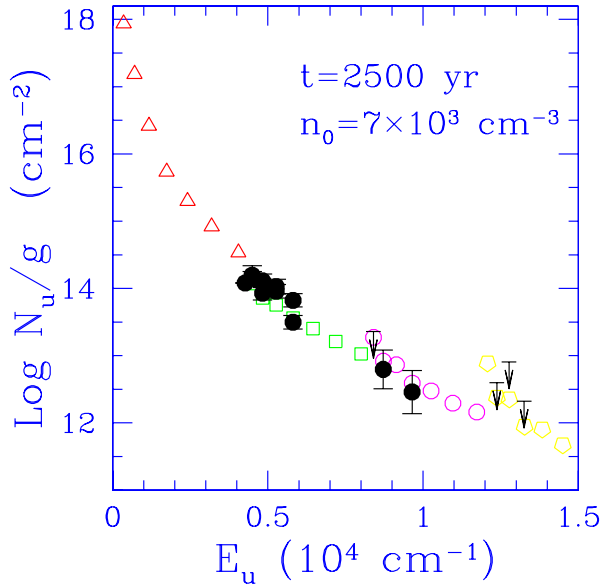}}
\caption{  Boltzmann plot.
The values of the column density of
the upper level of the transition (divided by its statistical weight)
are plotted as function  of the energy of the level.
The filled dots are the values derived from the observed lines.
Arrows are 3$\sigma$ upper limits.
The prediction of the best-fitting model ($n_0=7\times 10^3$ cm$^{-3}$) 
are shown
by the open triangles for lines of the v=0-0 band, open squares for
the 1-0, open circles for the 2-1 and pentagons for the 3-2 band.
}
\label{fig:h2col}
\end{figure}

The models  agree rather well
with the observations for all the lines. 
Fig.~\ref{fig:h2col} shows a Boltzmann plot for the \Hmol\
transitions, where the  column density of
the upper level of the transition (divided by its statistical weight)
is  plotted as function  of the energy of the level.
The filled dots are the  observed values, the open symbols
show the
prediction of the best-fitting model (Mod 1, $n_0=7\times 10^3$ cm$^{-3}$).
The agreement is somewhat worse
for models with higher density, which tend to predict weaker lines
from the higher excitation levels.

\subsection {Shock Models}

In the NH98 models of the \Hmol\ shock emission, 
the  relevant parameters that determine
the intensity of the lines
are the
shock velocity $v_{sh}$ (i.e.,  the difference between the
torus expansion  velocity and the red-giant wind velocity) and the pre-shock density,
which, at any given distance $R$ from the star,
is determined by the red-giant wind properties 
($n_w=\dot M_w/(4\pi R^2 \mu v_w f_w)$, where $\dot M_w$ is the rate of 
mass-loss, $v_w$ the wind velocity, $f_w$ the fraction of solid angle over which the wind was ejected
and $\mu$ the mean molecular weight). The \Hmol\ line intensity
does not depend on the torus properties, but only on its expansion velocity,
as long as the matter in the torus does not become completely ionized
or photodissociated.

\begin{figure}
\resizebox{\hsize}{!} {\includegraphics{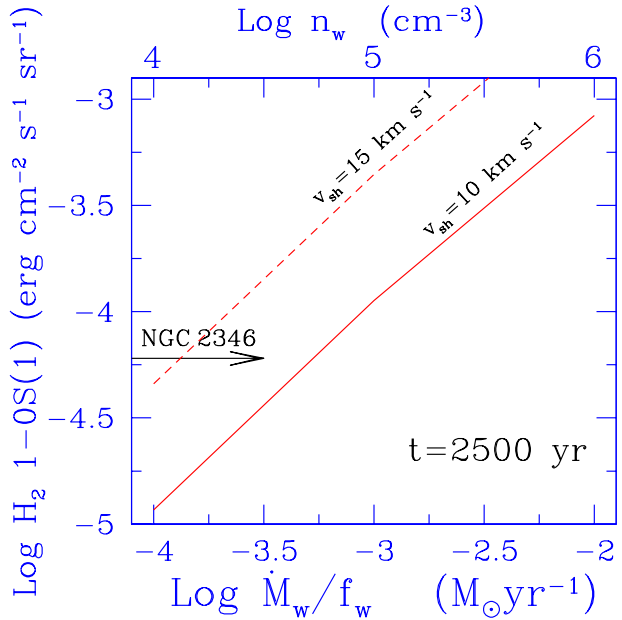}}
\caption{ Intensity of the 1-0S(1) line emitted in the shock between the
expanding torus and the precursor red-giant wind at 
$t=2500$ yr as function of
the red-giant wind parameter
$\dot M_w/f_w$ (see text).
The top scale gives the
corresponding value of the pre-shock density $n_w$ for $v_w=8$ \kms .
The two curves correspond to two values of
the shock velocity, as labelled.
The  observed 1-0 S(1) intensity  is shown by the horizontal arrow.
}
\label{fig:shock}
\end{figure}

In NGC~2346, the  expansion velocity of the neutral gas  derived from
CO data ranges from $\sim$ 15 to $\simgreat 35$ \kms
(Bachiller et al. 1989). We show in Fig.~\ref{fig:shock} the predicted
1-0S(1) intensity at $t=2500$ yr as a function of  the wind parameter
$\dot M_w/f_w$  and two shock velocities,
$v_{sh}$=10 \kms (solid line),
and $v_{sh}$=15 \kms (dashed line).
Higher values of $v_{sh}$ will appreciably dissociate H$_2$ in J shocks.
The intensity is
computed according to Eq.17
of NH98.
For $v_{sh}$=10 \kms,
one needs 
$\dot M_w/f_w\sim 4.4\times 10^{-4}$ \Myr 
in order to reproduce the  observed 1-0S(1) intensity. 
Unless $f_w\ll 0.1$,
this implies an unusually high  rate of mass-loss in the red-giant wind
(Loup et al. 1993). The density of the pre-shock gas ($n_w$)
is about $5\times 10^4$ \cmc; this value is also unconfortably
high, given the fact that the observed density of the  CO  torus 
(postshock gas), 
which  these
models predict to be $\gg n_w$,
is  $\simless 10^4$ \cmc\  (Bachiller et al. 1989).
A lower mass-loss rate is required if 
the shock velocity is higher. For $v_{sh}$=15 \kms,
the observed intensity is reproduced by $\dot M_w/f_w\sim 1.7\times 10^{-4}$
\Myr and a pre-shock density of about $2\times 10^4$ \cmc. 
The shock models of NH98 predict a ratio of the 2-1S(1)/1-0S(1) 
intensity of 0.10 for $v_{sh}=10$ \kms, and 0.19 for $v_{sh}=15$ \kms,
somewhat larger than the observed value ($\sim$0.07).

\section {Discussion}

The comparison of the observed intensity of the \Hmol\ 1-0S(1)
line in the central torus of NGC~2346 to the NH98 model
predictions shows that  the emission can be  produced in
the hot PDR generated by the radiation of the central star,
once the effect of X-ray heating and time-dependent (advecting)
chemistry
is taken into account. The best fit is obtained by models
with relatively low density of the neutral gas (in agreement
with the low density of the ionized material inferred by several
authors).  However, these PDR models require  the luminosity of the
hot central star to be significantly higher (\Lstar$\sim$250 \Lsun)
than current estimates.

If the \Hmol\ lines are emitted in the PDR, we expect
to observe a similar morphology in the ionized and \Hmol\
emitting gas.  In NGC~2346,
the PDR origin of the \Hmol\ lines 
is supported by the fact that the same morphology is seen
in \Hmol\ and in H$\alpha$ (see Walsh 1983). Also, we detect
\brg\ emission in the two \Hmol\ peaks, with a N-S profile that
follows that of the 1-0S(1) line (Fig.~\ref{fig:cuts}). The
intensity of  \brg\ predicted by the models is very low,
of the order of $8\times 10^{-6}$ \sbu, comparable
to the observed values (4--6$\times 10^{-6}$ \sbu).
This supports our estimate of \Lstar.

In principle, the observed intensity of the \Hmol\ 1-0S(1) line can also be 
accounted for by the emission of  the shocked gas produced by the
expansion of the torus inside a precursor red-giant wind.
However,
we estimate (following NH98) that one needs a rather high value
of the mass-loss rate in the red-giant wind
($\dot M_w/f_w\simgreat 10^{-4}$ \Myr). This, in turn,   implies a
high density of the pre-shock gas ($\simgreat 10^4$ \cmc), which is
not supported by any existing observation.
In fact, as discussed by
Zuckerman and Gatley (1988), the main difficulty in ascribing the
\Hmol\ vibrationally excited  emission to shocks comes
from the high momentum rate these models require.
Many authors (see the review by Kwok 1993) have proposed that
the formation and expansion of the PN
shell (or torus) is related to the action of the  fast wind from
the central star. In this case, the ambient gas (pre-shock
red giant wind) gains momentum approximately at the rate at which momentum
is delivered to the torus by the fast wind.
Since the fast wind is radiation driven, this rate
($\dot P$) must be $\simless$\Lstar/$c$.
In NGC~2346, we estimate that $\dot P$ is at least $\sim 2\times 10^{28}$ 
erg cm$^{-1}$ s$^{-2}$, i.e., more than 600 times the present value
of \Lstar/$c$ (for  \Lstar=250 \Lsun)
and 12 times higher than the maximum \Lstar/c reached by the
star in its earlier evolution, according to the evolutionary tracks of
Bl\"ocker (1995).

The interpretation of the \Hmol\ emission in terms
of shocks is often justified in the literature by the low 
measured ratio of the 2-1S(1) to the 1-0S(1)
intensity. This argument, however,  is not very strong, since in dense
PDRs the low vibrational \Hmol\ levels are thermalized.
The PDR models discussed in \S 4.1 predict a ratio 2-1S(1)/1-0S(1)$\sim$0.15
(not very different from the 0.10-0.18 range predicted by shock models;
see \S 4.2),
independently of the density and stellar luminosity.  These values are somewhat
higher than the observed ratio ($\sim$0.07).
It is possible, and worth further investigations, that models
tend to overestimate the fluorescent component of high
vibrational lines, possibly because of uncertainties in the
collisional deexcitation rates.

An interesting result of our observations, and one
that we cannot account for with our simple  models,
is  the variation
of the 2-1S(1)/1-0S(1)
ratio  with position along the slits, 
ranging from about 0.08 to 0.15 along the W slit, and from
0.08 to 0.23 along the E slit (see Fig.~\ref{fig:cuts}). These variations are
not monotonic with the distance from the peaks, but show evidence of structures,
especially along the W slit. It is possible that this is due to density
variations, which affect 
the fluorescent contribution  to the 2-1S(1) line. If the emission is due to
shocks, this could trace variations in the  propagation velocity of the
shock in an inhomogeneous medium.

Further support to the PDR origin of the \Hmol\ emission can be obtained by
observing lines from higher vibrational states. We show in Table 3
PDR model-predicted values for  lines
not detected so far in NGC~2346, which, however, are accessible from 
space.
The lines are very weak  with respect to the 1-0S(1), with ratios that do not 
depend significantly on the density. We expect that they will be 
substantially weaker
in shocks. 
To complete the discussion of the \Hmol\ spectrum, 
we show  in Table 4 model-predicted values of the intensity of
mid-infrared lines in the v=0-0 band for four of the PDR models 
described in \S 4.1.
These lines have been observed
by ISO in a number of PNe (among them NGC~2346; Barlow et al. in preparation),  
and may be useful diagnostic of the physical conditions (see NH98). 

\begin{table}
\begin{center}
\caption{ High-v H$_2$ PDR Predicted Line Ratios}
\vskip 0.1cm
\vbox{\hskip -8mm
\begin{tabular}{lccccc}
\hline\hline
\medskip
Line& $\lambda$& Mod 1& Mod 2& Mod 3& Mod5\\
& ($\mu$m) & \multispan4\hfil (1-0S(1)=100)\hfil \\
\hline
\smallskip
4-3S(1)& 2.5414&  15& 20&  20&  18\\
5-4S(1)& 2.7172&  0.6& 0.8& 0.8& 0.7\\
6-5S(1)& 2.9207&  0.2& 0.3& 0.3& 0.3\\
\hline
\hline
\end{tabular}}
\end{center}
\end{table}

\begin{table}
\begin{center}
\caption{ Mid-Infrared H$_2$ PDR Predicted Line Intensities}
\vskip 0.1cm
\vbox{\hskip -8mm
\begin{tabular}{lccccc}
\hline\hline
\medskip
Line& $\lambda$& Mod 1& Mod 2& Mod 3& Mod5\\
& ($\mu$m) & \multispan4\hfil (10$^{-5}$erg cm$^{-2}$ s$^{-1}$ sr$^{-1}$)\hfil \\
\hline
\smallskip
0-0S(0) & 28.22& 0.07& 0.06& 0.03& 0.05 \\
0-0S(1) & 17.03& 1.4 & 1.3 & 0.6 & 1.2 \\
0-0S(2) & 12.28& 0.8 & 1.0 & 0.6 & 0.7 \\
0-0S(3) &  9.66& 2.8 & 3.1 & 2.5 & 1.9 \\
0-0S(4) &  8.02& 1.3 & 1.0 & 0.9 & 0.5 \\
0-0S(5) &  6.90& 5.0 & 2.7 & 1.9 & 1.0 \\
0-0S(6) &  6.11& 1.7 & 0.7 & 0.4 & 0.3 \\
\hline\hline
\end{tabular}}
\end{center}
\end{table}


\section {Summary and Conclusions}

This paper is motivated by our interest in understanding
the origin of the \Hmol\ emission in PNe.  Recently,
Natta \& Hollenbach (1998) have computed the evolution with time
of the \Hmol\ emission expected in PNe of different core mass and
shell (or torus) properties. Their models compute the emission originating
in the photodissociation region (PDR) created by the UV radiation 
of  the central core incident on the inner edge of the neutral shell, as well as 
the emission in the shock associated to the expansion of the torus
inside the precursor red-giant wind. NH98 show that both regions
can produce intense \Hmol\ emission  and that detailed studies
of individual PNe are necessary.

NGC~2346 is a good candidate. It is a bright, young PN, with a typical
butterfly morphology characterised by well developed
bipolar lobes of emission and a bright torus around the central star.
The same morphology is seen in H$\alpha$, CO and in the \Hmol\ 1-0S(1)
line. There is a large amount of information in the literature,
concerning the properties of the ionized region and the gas kinematic  
that can be used in our analysis.

We have collected  near infrared  observations
of NGC~2346. The data
include broad K band image, an image in the \Hmol\ vibrationally excited
1-0S(1) line  and slit spectra
in the K band in three positions in the nebula.
The images 
confirm
the well-known NGC~2346 morphology, with a central,
bright torus, surrounded by weaker emission with a typical
butterfly shape. 
The K band spectra show 11 \Hmol\ lines with excitation energies from
6150 to 12552 K.
Profiles of the lines intensity along the slit show evidence of
secondary condensations outside  the torus midplane.

A comparison of our observations with the NH98 models shows that
PDR emission can account for the  \Hmol\ observations. This requires
a low-density shell ($n\simless 10^4$ cm$^{-3}$), in agreement with the low
density measured in the ionized region. Note that steady-state models or
models where the soft X-ray radiation from the star is ignored predict
a 1-0S(1) intensity one order of magnitude lower 
than models where both these effects are included.
PDR models of the \Hmol\ emission needs a central star significantly
more luminous than estimated in the literature (250 \Lsun\ against
17--90 \Lsun). However,
we think that this is consistent with all the available data,
including our own \brg\ observations.

It is unlikely that the
\Hmol\ emission originates in the shock between the expanding shell
and the precursor red giant wind.
Shock models require
a much larger momentum input to the torus than possible from the central
star. We estimate the discrepancy to be at least a factor 600 for
the current luminosity of the central star (taken to be 250 \Lsun)
and a factor 12 if we consider the maximum luminosity ever reached by
the star in its previous evolution.

In conclusion, we have proved that the PDR origin for the \Hmol\
emission in NGC~2346 is  likely and that the models, even if
very simple, can account for a number of the observed properties.
We suggest some additional observations of 
lines from higher excitation vibrational  levels and 
in the 0-0 band, which may help in determining the physical conditions
in the shell with higher accuracy.

\begin{acknowledgements}
We thank Leonardo Vanzi for his help during observations,
Sandro Gennari and the LONGSP/TIRGO staff for their help in this project.
We are indebted to
Frank Bertoldi, Tino Oliva  and Malcolm Walmsley
for many interesting discussions on the various aspects of this work.
The theoretical project on time-dependent 
PDR  was partly supported by ASI grants 92-RS-54, 
94-RS-152,  ARS-96-66 
to the Osservatorio di Arcetri.
Support from C.N.R.--N.A.T.O. Advanced Fellowship program and
from NASA's {\it Origins of Solar Systems} program (through grant NAGW--4030)
is gratefully acknowledged.

\end{acknowledgements}

\end{document}